\documentclass[10pt]{article}

\usepackage{latexsym,amsmath,amssymb,epsfig}

\DeclareFontFamily{OT1}{rsfs10}{}
\DeclareFontShape{OT1}{rsfs10}{m}{n}{ <-> rsfs10 }{}
\DeclareMathAlphabet{\mathscript}{OT1}{rsfs10}{m}{n}

\numberwithin{equation}{section}

\newcommand{\be}{\begin{equation}}
\newcommand{\ee}{\end{equation}}
\newcommand{\nn}{\nonumber}
\newcommand{\bea}{\begin{eqnarray}}
\newcommand{\eea}{\end{eqnarray}}

\newcommand{\ns}{\normalsize}
\newcommand{\pt}{\partial}

\def\a{\alpha}
\def\b{\beta}
\def\g{\gamma}

\def\d{\delta}
\def\e{\epsilon}

\def\f{\phi}

\def\k{\kappa}

\def\n{\nu}

\def\s{\sigma}
\def\t{\tau}

\def\x{\xi}

\def\w{\wedge}

\def\G{\Gamma}

\def\cA{{\cal A}}

\def\cK{{\cal K}}
\def\cC{{\cal C}}
\def\cP{{\cal P}}

\def\cV{{\cal V}}
\def\cW{{\cal W}}

\begin{document}


\begin{titlepage}

\vspace{-3cm}

\title{
   \hfill{\ns SUSX-TH/02-027\\}
   \hfill{\ns HU-EP-02/52\\}
   \hfill{\ns hep-th/0212263}\\[3em]
   {\huge Flop Transitions in M-theory Cosmology}\\[1em]}
   \setcounter{footnote}{0}
\author{
{\ns\large Matthias Br\"andle$^1$\footnote{email: brand@physik.hu-berlin.de}
  \setcounter{footnote}{3}
  and Andr\'e Lukas$^2$\footnote{email: a.lukas@sussex.ac.uk}} \\[0.8em]
   {\it\ns $^1$Institut f\"ur Physik, Humboldt Universit\"at}\\[-0.2em]
   {\ns Invalidenstra\ss{}e 110, 10115 Berlin, Germany}\\[0.4em]
   {\it\ns $^2$Centre for Theoretical Physics, University of Sussex}
   \\[-0.2em]
   {\ns Falmer, Brighton BN1 9QJ, UK} \\[0.2em] }
\date{}

\maketitle

\begin{abstract}
We study flop-transitions for M-theory on Calabi-Yau three-folds
and their applications to cosmology in the context of the
effective five-dimensional supergravity theory. In particular,
the additional hypermultiplet which becomes massless at the
transition is included in the effective action. We find
the potential for this hypermultiplet which includes quadratic
and quartic terms as well as additional dependence on the K\"ahler
moduli. By constructing explicit cosmological solutions, it is
demonstrated that a flop-transition can indeed by achieved
dynamically, as long as the hypermultiplet is set to zero.
Once excitations of the hypermultiplet are taken into account
we find that the transition is generically not completed but
the system is stabilised close to the transition region.
Regions of moduli space close to flop-transitions can, therefore,
be viewed as preferred by the cosmological evolution.
\end{abstract}

\thispagestyle{empty}

\end{titlepage}


\section{Introduction}

Space-like curvature singularities arising in cosmological solutions
to low-energy string effective actions and their potential resolution
constitute a challenging problem in string and M-theory. On the other
hand, the string resolution of certain time-like singularities, such
as those arising from collapsed cycles in the internal manifold, is,
at least in principle, understood. In the course of a string/M-theory
phase transition, triggered by cosmological evolution of moduli fields,
these singularities may, in fact, arise at a particular instance in
time. For example, a
flop-transition~\cite{Aspinwall:1993nu,Witten:1996qb,Greene:1996cy}
corresponds to a collapsing two-cycle in the internal Calabi-Yau space
while a conifold
transition~\cite{Strominger:1995cz,Greene:1995hu,Greene:1996cy}
corresponds to a collapsing three-cycle. Clearly, such transitions are
of interest for string and M-theory early universe cosmology. For
example, one would like to know whether the topological transition can
actually be realized dynamically, that is, whether the topology of the
internal manifold could have indeed changed during the cosmological
evolution. Further, one would like to understand, in this cosmological
context, the role of the states which become light at the transition
and how precisely the transition effects the evolution of the fields.

In this paper, we will be answering these questions for the mildest
form of topology change, namely the flop. A related discussion, but in
the context of black-hole solutions, has been carried out in
Ref.~\cite{Chou:1997ba,Gaida:1998km}. We will be working in the
context of M-theory on Calabi-Yau three-folds leading to an effective
description in terms of five-dimensional $N=1$ supergravity
theories~\cite{Chamseddine:1980sp}--\cite{Ceresole:2000jd}. Flop-transitions
arise from collapsing two-cycles within the Calabi-Yau manifold and
are, therefore, controlled by the K\"ahler moduli which, together with
$U(1)$ gauge fields, are contained in five-dimensional vector
multiplets. Membranes wrapping Calabi-Yau two-cycles lead to
hypermultiplet states in five dimensions with a mass proportional to
the volume of the cycle. When the cycle collapses at the transition
the hypermultiplet becomes massless and can no longer be ignored in
the effective theory. In our analysis, we will include these
hypermultiplet states explicitly into the five-dimensional
effective action. In the following, we will also refer to these states
as "transition states". In the case of M-theory on Calabi-Yau three-folds,
there are no non-geometrical phases~\cite{Witten:1996qb}, that is,
transitions are sharp. This implies that, going through the transition
by first collapsing the cycle and then blowing it up in a
topologically different way, leads to a another, topologically
distinct Calabi-Yau space. While the Hodge numbers of the original and
the ``flopped'' Calabi-Yau space are the same other topological
quantities, such as the intersection numbers, change across the
transition.

In terms of the five-dimensional effective supergravity theory, the
transition can be described in a non-singular way once the additional
hypermultiplet is included. For example, the jump in the intersection
numbers which appear in the five-dimensional Chern-Simons term is
accounted for by loop-corrections involving the hypermultiplet
states~\cite{Witten:1996qb} while the K\"ahler moduli space metric is continuous
across the transition~\cite{Chou:1997ba}. It turns out that the additional
hypermultiplet is charged with respect to a particular linear
combination of the vector multiplet gauge fields. Supersymmetry then
implies the existence of a potential which depends on the transition states
and the vector multiplet scalars. It is this potential which will play
an important role in our cosmological analysis. Practically, we will,
therefore, study time-evolution in K\"ahler moduli space close to the
flop region including the effect of the transition states and their
potential.

The plan of the paper is as follows. In the next section, we will
review $N=1$ supergravity in eleven and five dimensions and the
five-dimensional effective action for M-theory on Calabi-Yau
three-folds. With this machinery at hand, we then go on to derive the
effective five-dimensional action of the transition states.  Section 3
analyses the cosmology of the five-dimensional theory for arbitrary
Calabi-Yau spaces, first with vanishing and then with non-vanishing
transition states. In section 4, we focus on a specific example of two
Calabi-Yau spaces related by a flop and study the cosmological
evolution numerically. We conclude in section 5.


\section{The five-dimensional effective action of M-theory}

To set the notation, we will first review $N=1$ supergravity in eleven
and five dimensions and the structure of the five-dimensional
effective action for M-theory on Calabi-Yau three-folds. Subsequently,
we will show how to couple to this action the hypermultiplet
which contains the transition states. The five-dimensional effective
action including this hypermultiplet will be the basis for the
subsequent cosmological analysis.

\subsection{Supergravity in eleven and five dimensions}

The bosonic part of eleven-dimensional supergravity is given
by~\cite{Cremmer:1978km}
\begin{equation}
 S_{11}  = -\frac{1}{2\k^{2}}\int_{M_{11}}\left\{d^{11}x\,
        \sqrt{-g}\,\left(\frac{1}{2}R+\frac{1}{4!}G_{IJKL}G^{IJKL}\right)
                +\frac{2}{3}C\w G\w G\right\}\,,\\ \label{S11}
\end{equation}
where $G=dC$ is the field strength of the three-form potential $C$ and
$\k$ is the 11-dimensional Newton constant, as usual. Indices
$I,J,K,\dots = 0,\dots ,10$ label the 11-dimensional coordinates $x^I$.
Later, we will also need the bosonic part of the membrane
action~\cite{Howe:hp,Bergshoeff:1987cm}
\begin{equation}
   S_{M_{3}} = -T_{2}\int_{M_{3}}\left\{d^{3}\s\sqrt{-\g}+
     2\hat{C}\right\}\, \label{SM3}
\end{equation}
which couples to $C$. The membrane world-volume is parametrised by
coordinates $\s^n$, where $n,m,p,\dots =0,1,2$, and its embedding into
11-dimensional space-time is specified by $X^I=X^I(\s )$. The
pull-backs $\g_{nm}$ and $\hat{C}$ of the space-time metric and the
three-form are defined by
\begin{eqnarray}
  \g_{nm}&=&\pt_{n}X^{I}\pt_{m}X^{J}g_{IJ}\, ,\\
  \hat{C}_{nmp}&=&\pt_{n}X^{I}\pt_{m}X^{J}\pt_{p}X^{K}C_{IJK}\, ,
\end{eqnarray}
as usual. In terms of the 11-dimensional Newton constant, the membrane
tension $T_2$ is given by
\begin{equation}
    T_{2}=\left(\frac{8\pi}{\k^{2}}\right)^{\frac{1}{3}}\, .
\end{equation}

\vspace{0.4cm}

Let us now move on to five-dimensional $N=1$ supergravity
focusing on the aspects relevant to this paper. For a more complete
account we refer to the
literature~\cite{Chamseddine:1980sp}--\cite{Ceresole:2000jd}.

We denote five-dimensional space-time indices by $\a ,\b ,\g ,\dots =
0,1,2,3,4$. In addition to the supergravity multiplet, consisting of
the vielbein, an Abelian vector field and the gravitini, there are two
types of matter multiplets, namely vector- and hyper-multiplets. In
general, one can have any number, $n_{\rm V}$, of vector multiplets each
containing a real scalar field and an Abelian vector field plus
fermionic partners and any number, $n_{\rm H}$, of hypermultiplets each
containing four real scalars plus fermions. It is useful to treat the
Abelian gauge fields in the vector multiplets and the supergravity
multiplet on the same footing and collectively denote them by $A^i_\a$
where $i,j,k,\dots =0,\dots ,n_{\rm V}$. The real scalars
contained in the vector multiplets are described by $n_{\rm V}+1$
fields $b^i$. These define a manifold of very special
geometry~\cite{Gunaydin:1983bi} with metric
\begin{equation}
 G_{ij} = -\pt_i\pt_j\ln K\; , \label{G}
\end{equation}
which is given in terms of the degree three homogeneous polynomial
\begin{equation}
 K = d_{ijk}b^ib^jb^k\, .\label{K}
\end{equation}
Here $d_{ijk}$ are constant coefficients. The $b^i$ are subject
to the constraint
\begin{equation}
 K=6 \label{cons}
\end{equation}
which reduces the number of independent fields to $n_{\rm V}$, as required.

Further, we denote by $Q^u$, where $u,v,w,\dots = 1,\dots ,4n_{\rm
H}$, the hypermultiplet scalars. They parametrise a quaternionic
manifold, that is, a manifold with holonomy $SU(2)\times Sp(2n_{\rm
H})$. The metric on this manifold, $h_{uv}$, is hermitian with respect
to the three complex structures
\begin{equation}
 {J_u}^v={J_u^a}^v\t_a
\end{equation}
satisfying the quaternionic algebra
\begin{equation}\label{q-algebra}
  J^{a\;\;v}_{\;u}J^{b\;\;w}_{\;v}=
  -\d^{ab}\d_{u}^{\;w}+\e^{abc}J^{c\;\;w}_{\;u}\,,
\end{equation}
where $a,b,c,\dots =1,2,3$. Here $\t_a$ are the hermitian Pauli matrices
so that the complex structures fill out the adjoint of $SU(2)$. The
associated triplet of K\"ahler forms is given by
\begin{equation}
 \cK_{uv}= {J_u}^wh_{wv}\, .
\end{equation}
We also need to introduce the $SU(2)$ part $\omega_u=\omega_u^a\t_a$ of the
spin connection.

Let us assume that the metric $h_{uv}$ admits $n_{\rm V}+1$ Killing
vectors $k_i^u$. These Killing vectors should respect the quaternionic
structure which means they originate from prepotentials
$\cP_i=\cP_i^a\t_a$ via the relation
\begin{equation}\label{defprepot}
   k_{i}^{u}\cK_{uv}=\pt_{v}\cP_{i}+[\omega_{v},\cP_{i}]\, .
\end{equation}

With these conventions the bosonic part of the supergravity and vector
multiplet action reads
\begin{equation}
      S_{\rm V}=-\frac{1}{2\k_{5}^{2}}\int_{M_{5}}\bigg\{d^{5}x\sqrt{-g}
            \bigg(\frac{1}{2}R+\frac{1}{4}G_{kl}\pt_{\a}b^{k}\pt^{\a}
             b^{l} +\frac{1}{2}G_{kl}F^{k}_{\a\b}F^{l\,\a\b}\bigg)
             +\frac{2}{3}d_{klm}\,A^{k}\w F^{l}\w F^{m}\bigg\}\,,
      \label{SV}
\end{equation}
where $\k_5$ is the five-dimensional Newton constant.
The bosonic part of the hypermultiplet action takes the form
\begin{equation}
S_{\rm H}=-\frac{1}{2\k_{5}^{2}}\int_{M_{5}}d^{5}x\sqrt{-g}\left\{
   h_{uv}D_{\a}Q^{u}D^{\a}Q^{v}+V\right\}\, ,\label{SH}
\end{equation}
with the potential $V$ given by
\begin{equation}
V = \frac{1}{2}g^{2}\left[4(G^{ij}- b^{i}b^{j})\mbox{tr}(\cP_{i}\cP_{j})
         +\frac{1}{2}b^{i}b^{j}h_{uv}k_{i}^{u}k_{j}^{v}\right]\, .
\label{V}
\end{equation}
The trace in this expression is performed over the Pauli matrices.
The covariant derivative $D_\a$ includes the gauging of the hypermultiplets
with respect to the vector fields $A^i$ and is defined by
\begin{equation}\label{covder}
  D_{\a}Q^{u}=\pt_{\a}Q^{u}+g\,A^{i}k_{i}^{u}\,,
\end{equation}
where $g$ is the gauge coupling. Note that the appearance and the structure
of the potential~\eqref{V} is directly linked to this gauging of the
hypermultiplets.

\subsection{M-theory on Calabi-Yau three-folds}

Let us now briefly review the reduction~\cite{Cadavid:1995bk} of the
action~\eqref{S11} for 11-dimensional supergravity on a Calabi-Yau
three-fold $X$ with Hodge numbers $h^{1,1}$ and $h^{2,1}$. This leads
to a five-dimensional $N=1$ supergravity theory of the type described
in the previous subsection with $n_{\rm V}=h^{1,1}-1$ vector
multiplets, $n_{\rm H}=h^{2,1}+1$ hypermultiplets and no gauging of
the hypermultiplets.

We now need to specify the geometrical origin of some of these 
five-dimensional fields. The 11-dimensional metric on the direct
product space $M_{11}=M_5\times X$ can be written as
\begin{equation}
  ds_{11}^{2}=\cV^{-2/3}g_{\a\b}dx^{\a}dx^{\b}+g_{AB}dx^{A}dx^{B}\,,
\end{equation}
where $A,B,C,\dots = 5,\dots ,10$ label the coordinates on the
Calabi-Yau space, $g_{AB}$ is the Ricci-flat metric on $X$ and
$g_{\a\b}$ is the five-dimensional metric. The Calabi-Yau volume
modulus $\cV$ is defined by
\begin{equation}
  \cV =\frac{1}{v}\int_{X}d^{6}x\sqrt{g_{6}}\, ,
\end{equation}
where $v$ is an arbitrary six-dimensional reference volume. We note
that the five-dimensional Newton constant $\k_5$ is related to its
11-dimensional counterpart $\k$ by
\begin{equation}
 \k_5^2 = \frac{\k^2}{v}\, .
\end{equation}
The K\"ahler form $\omega_{AB}$ of $X$ can be expanded as
\begin{equation}
 \omega_{AB} = a^i\omega_{iAB}
\end{equation}
into a basis $\{\omega_{iAB}\}$ of $(1,1)$ forms. We take this basis of $(1,1)$
forms to be dual to an (effective) basis $\{\cW^i\}$ of the second homology,
that is,
\begin{equation}
v^{-1/3}\int_{\cW^i}\omega_j = \d_j^i\; .
\end{equation}
The $h^{1,1}$ expansion coefficients $a^i$ are the K\"ahler moduli of the
Calabi-Yau space. As stands, they are, of course, not independent from
the volume modulus $\cV$. However, we can define volume-independent
moduli $b^i$ by
\begin{equation}
 b^i = \cV^{-1/3}a^i\; .
\end{equation}
These constitute $n_{\rm V}=h^{1,1}-1$ independent fields as they
can be shown to satisfy the constraint~\eqref{cons}. They
should be interpreted as the scalar fields in the vector multiplets.
The coefficients $d_{ijk}$ which appear in Eq.~\eqref{K} should then
be identified as the intersection numbers of the Calabi-Yau space.

The three-form $C$ can be expanded as
\begin{equation}
   C=\bar{C}+A^{i}\w\omega_{i}+\x\w\Omega+\bar{\x}\w\bar{\Omega}\,,\label{C}
\end{equation}
where $\Omega$ is the holomorphic $(3,0)$ form on $X$. The $h^{1,1}$
five-dimensional vector fields $A^i_\a$ account for the gauge
fields in the vector multiplets and in the gravity multiplet. The
five-dimensional three-form $\bar{C}$ can be dualised to a scalar and
forms, together with the complex scalar $\x$ and the volume modulus
$\cV$, the universal hypermultiplet. There are $h^{2,1}$ additional
hypermultiplets which originate from the complex structure moduli and
the $(2,1)$ part of $C$ which we have omitted in Eq.~\eqref{C}.
These standard hypermultiplets will not be of particular importance, in the
following.

\vspace{0.4cm}

Let us now review some relevant features of M-theory flop transitions
following Refs.~\cite{Witten:1996qb,Chou:1997ba}. A flop constitutes a
transition from the Calabi-Yau space $X$ to a topologically different
space $\tilde{X}$ due to a complex curve $\cC$ in $X$ shrinking to
zero size and subsequently being blown up in a topologically distinct
way. Concretely, let us expand the class $\cW$ of the curve $\cC$
in our homology basis
as
\begin{equation}
 \cW = \b_i\cW^i
\end{equation}
with constant coefficients $\b^i$. The volume of $\cC$ can then be
written as
\begin{equation}
 {\rm Vol}(\cC ) = \int_{\cC}\omega = (v\cV )^{1/3}b
\end{equation}
where we have introduced the particular linear combination
\begin{equation}
 b = \b_ib^i \label{b}
\end{equation}
of vector multiplet moduli. Within the moduli space of $X$ we have
$b^i>0$, for all $i$, as well as $b>0$ and the limit $b\rightarrow 0$
corresponds to approaching the flop. Continuing further to negative
values of $b$ leads into the moduli space of the birationally
equivalent Calabi-Yau space $\tilde{X}$. This new Calabi-Yau space
$\tilde{X}$ has the same Hodge numbers and, hence, the same
five-dimensional low-energy spectrum as the original space
$X$. However, the intersection numbers have changed across the
transition~\cite{Witten:1996qb}. More specifically, setting
$(\b_i)=(1,0,\dots ,0)$ for simplicity the new intersection numbers
$\tilde{d}_{ijk}$ (expressed in terms of the field basis $b^i$) are
given by
\begin{equation}
 \tilde{d}_{111}=d_{111}-\frac{1}{6} \label{ichange}
\end{equation}
with all other components unchanged. Sometimes a new basis of fields
$\tilde{b}^i$, defined by
\begin{equation}
 \tilde{b}^1=-b^1\; , \qquad \tilde{b}^i=b^i-b^1\; , \label{bt}
\end{equation}
for all $i\neq 1$, is introduced~\cite{Chou:1997ba} to cover the
moduli space of $\tilde{X}$. These new fields have the advantage of
being positive throughout the moduli space of $\tilde{X}$ which is not
the case for the original fields $b^i$. For our applications we will
find it usually more practical to use a single set of fields to cover 
the moduli spaces for both $X$ and $\tilde{X}$.

How does the five-dimensional effective theory change across the
transition? Inspection of the action~\eqref{SV}, \eqref{SH}
without gauging and potential shows that only the vector multiplet
part~\eqref{SV} is affected through the change~\eqref{ichange}
in the intersection numbers. From Eq.~\eqref{G}, the metric $G_{ij}$
takes the specific form
\begin{equation}
 G_{ij} = -d_{ijk}b^k+\frac{1}{4}d_{ikl}d_{jmn}b^kb^lb^mb^n\; ,
\end{equation} 
where we have used that $K=6$. This form shows that, despite the
jump~\eqref{ichange} in the intersection number the metric remains
continuous across the flop since $d_{111}$ is always multiplied by $b^1$
which vanishes at the transition. We remark that the associated connection
\begin{equation}
 \G_{ij}^k = \frac{1}{2}G^{kl}\frac{\pt G_{ij}}{\pt b^l}
 \label{Gamma}
\end{equation}
which appears in the five-dimensional equations of motion contains a
term proportional to $d_{ijk}$ (without additional fields $b^i$) and,
hence, jumps across the flop. Given the continuity of the metric
$G_{ij}$ the only discontinuous term in the action is the Chern-Simons
term in Eq.~\eqref{SV} which is proportional to the intersection
numbers. It has been shown~\cite{Witten:1996qb}, that its jump can be
accounted for by loop corrections which involve the transition states.
Let us now discuss these additional states in more detail.

\subsection{The transition states}

The five-dimensions particles which become massless at the flop
originate from a membrane which wraps the collapsing complex curve $\cC$
with homology class $\cW$. We can find the world-line action for these
transition states by starting with the membrane action~\eqref{SM3}.
Introducing a complex world-volume coordinate $\s = \s^1+i\s^2$ and
world-time $\t = \s^0$ we consider an embedding of the membrane into
11-dimensional space of the form
\begin{equation} 
 X^\a = X^\a(\t )\; ,\qquad
 X^A = X^A(\s )\; ,\qquad
 X^{\bar{A}} = X^{\bar{A}}(\bar{\s})\; ,
\end{equation}
where here $A$ and $\bar{A}$ are holomorphic and anti-holomorphic
indices on the Calabi-Yau space, respectively, and $X^A=X^A(\s )$
parametrises the complex curve $\cC$. The reduction of the membrane
action on this curve leads to the following world-line action
\begin{equation}
  S_{p}=-(v^{1/3}T_{2})\int_{\mathbb{R}}\left\{dt\,(\b_{i}b^{i})
    \sqrt{-\pt_{\t}X^{\a}\pt_{\t}X^{\b}g_{\a\b}}+
    2\,\b_{i}\hat{A}^{i}\right\}\, . \label{Sp}
\end{equation}
This particle has four transverse (scalar) degrees of freedom and
must, hence, form a hypermultiplet in five dimensions. We denote the
scalars in this hypermultiplet by $q^u$, where $u,v,w,\dots =
1,2,3,4$. It is charged with respect to the particular linear
combination
\begin{equation}
    \cA\equiv\b_{i}A^{i}\,, \label{A}
\end{equation}
of vector fields with associated gauge coupling
\begin{equation}
  g= 2\,v^{1/3}T_{2}=
  2\left(\frac{8\pi}{\k_{5}^{2}}\right)^{1/3}\, ,\label{g}
\end{equation}
as can be seen from the last term in~\eqref{Sp}. From the first term in
the world-line action we can read off the mass which is given by 
\begin{equation}
  m=\frac{1}{2}g b =T_{2}\cV^{-1/3}\mbox{Vol}(\cC)\,. \label{m}
\end{equation}
What does this information tell us about the five-dimensional effective
action of these transition states? Clearly, these states being
hypermultiplets, their effective action must be of the general
form~\eqref{SH}. We assume that the associated hypermultiplet moduli
space metric is flat, so that
\begin{equation}
 h_{uv}=\d_{uv}\, . \label{h}
\end{equation}
As we will see shortly, this assumption is consistent with the
above properties of the transition states and the constraints enforced by
five-dimensional supergravity. To work this out explicitly, let us first
recall the quaternionic structure on the four-dimensional flat 
moduli space. Introducing the t'Hooft $\eta$-symbols~\cite{'tHooft:fv}
\begin{eqnarray}
 \eta^{a}_{bc}=&\bar{\eta}^{a}_{bc}&=\e^{a}_{bc}\,,\\
 \eta^{a}_{b0}=&\bar{\eta}^{a}_{0b}&=\d^{a}_{b}\,,
\end{eqnarray}
which satisfy the properties
\begin{eqnarray}\label{etaprops}
  [\eta^{i},\bar{\eta}^{j}]&=&0\,, \\
  (\eta^{i})^{T}=(\eta^{i})^{-1}\,,&&
  (\bar{\eta}^{i})^{T}=(\bar{\eta}^{i})^{-1}\, ,\nn
\end{eqnarray}
the triplet of complex structures can be written as
\begin{equation}
  J^{a\;\;v}_{\;u}\equiv-\bar{\eta}^{a}_{uw}\d^{wv}\,,
\end{equation}
which satisfy the quaternionic algebra~\eqref{q-algebra}, as required.
The associated triplet of K\"ahler forms is given by
\begin{equation}\label{K-forms}
  \cK^{a}_{uv}=-\bar{\eta}^{a}_{uv}.
\end{equation}
We know that the transition states are charged under the particular
combination of gauge fields~\eqref{A}. Hence the vectors $k_i^u$ must
be proportional to $\b_i$ and, at the same time, be Killing vectors on
flat four-dimensional space. We know that this gauging must lead to a
potential of the form~\eqref{V}. As $q^u\rightarrow 0$ this potential
must vanish so that the moduli $b^i$ indeed parametrise flat
directions in this limit. This implies that the Killing vectors
$k_i^u$ should not correspond to translations but rather to rotations
and, hence, be of the form
\begin{equation}
 k_i^u = \b_i{t^u}_vq^v\, ,\label{k}
\end{equation}
where $t$ is an arbitrary anti-symmetric matrix. In addition,
these Killing vectors must originate from a prepotential, that is,
they must satisfy Eq.~\eqref{defprepot}. This is the case precisely
if $[t,\bar{\eta}^a]=0$ for $a=1,2,3$, or, equivalently, if the
matrix $t$ is of the form
\begin{equation}
 t_{uv} = n_a\eta^a_{uv}\; ,\label{t}
\end{equation}
where $n_a$ are real coefficients. This matrix represents the
generator of $SO(2)$ in the representation ${\bf 2}\oplus{\bf 2}$.
We require the standard normalisation ${\rm tr}(t^2)=-4$
or, equivalently, $n_an^a=1$. The associated prepotential
then reads
\begin{equation}\label{prepots}
  \cP^{a}_{i}=\frac{1}{2}\beta_{i}q^{v}(\bar{\eta}^{a}_{vw}\,n_{b}
  \eta^{b\,w}_{\quad u})q^{u}+\x_{i}^{a}\,, 
\end{equation}
where $\x_i^a$ are arbitrary integration constant. They represent the
generalisation of Fayet-Illiopoulos terms to five-dimensional $N=1$
supergravity. As they lead to terms in the potential which do not
vanish for vanishing $q^u$ we will set them to zero in the following.

Inserting~\eqref{h}, \eqref{k}, \eqref{t} and \eqref{prepots} into
the general hypermultiplet action~\eqref{SH} we obtain
\begin{equation}
S_{q}=-\frac{1}{2\k_{5}^{2}}\int_{M_{5}}d^{5}x\sqrt{-g}\left\{
   D_{\a}q^{u}D^{\a}q_{u}+V)\right\}\,, \label{SHspec}
\end{equation}
with the potential
\begin{equation}
  V=\frac{1}{4}g^{2}\left[b^{2}q_{u}q^{u}
  +4(G^{kl}\b_{k}\b_{l}-b^{2})(q_{u}q^{u})^{2}\right]\label{Vspec}
\end{equation}
and the covariant derivative
\begin{equation}
 D_\a q^u = \pt_\a q^u +g\cA t_{uv}q^v\; . \label{Dspec}
\end{equation}
Consequently, the hypermultiplet current $j_{\a}$ which couples
to the gauge field $\cA_\a$ is given by
\begin{equation}
 j_{\a} = gq^ut_{uv}\partial_\a q^v\; .\label{current}
\end{equation}
We recall that $b=\b_ib^i$, defined in Eq.~\eqref{b}, is proportional
to the volume of the collapsing cycle and the generator $t$ has been
given in Eq.~\eqref{t}. So far, we have only used that the transition
states are charged under the gauge field $\cA$. Clearly, the gauge
coupling $g$ which appears in the covariant derivative~\eqref{Dspec}
has to be identified with the value~\eqref{g} obtained from the
reduction of the membrane action. Then, the above hypermultiplet
action~\eqref{SHspec} is completely fixed. From the first term in the
potential~\eqref{Vspec} we can now read off the mass of the
hypermultiplet which is given by $gb/2$. This value indeed coincides
with the one obtained from the membrane reduction, Eq.~\eqref{m}, as
it should for consistency.

In addition, we have found a potential term quartic in the transition
states which was not anticipated from the membrane reduction but imposed
on us by five-dimensional supergravity. This quartic term plays
an important role in lifting ``unwanted'' flat directions.
While the potential should be flat for vanishing
transition states, $q^u=0$, and arbitrary $b^i$, a flat direction
along the flop at $b=0$ and arbitrary $q^u$ would be a
surprise~\footnote{Such a flat direction with non-vanishing
transition states would correspond to a Higgs branch where the gauge symmetry
corresponding to the vector field $\cA$ is broken. The resulting change in
the number of light vector multiplets would be inconsistent with the
fact that Hodge numbers are unchanged across the flop.}.
Fortunately, this potential flat direction is lifted by the
second term in Eq.~\eqref{Vspec}.


\section{Cosmology}

\subsection{Cosmological ansatz and equations of motion}

Let us briefly summarise the discussion so far. We have seen that
M-theory on a Calabi-Yau three-fold $X$ with Hodge numbers $h^{1,1}$
and $h^{2,1}$ is effectively described by the five-dimensional
supergravity action~\eqref{SV}, \eqref{SH} with $n_{\rm V}=h^{1,1}-1$
vector multiplets and $n_{\rm H}=h^{2,1}+1$ hypermultiplets and no
gauging. When a flop-transition to a topologically distinct Calabi-Yau
space $\tilde{X}$ occurs the Hodge numbers and hence the number of
massless particles remains the same while the structure of the
five-dimensional action changes in accordance with the
change~\eqref{ichange} in the intersection numbers. In addition, at
and near the flop-transition region another light hypermultiplet
appears whose action~\eqref{SHspec} has to be added to the previous
one for an accurate description across the transition. It is important
not to confuse these transition hypermultiplet states which arises at
the flop with the standard hypermultiplets associated with the complex
structure moduli space of the Calabi-Yau space.

\vspace{0.4cm}

Which parts of this five-dimensional effective action are we actually
interested in for our cosmological applications? Since we would like
to study flop-transitions which arise by moving in the Calabi-Yau
K\"ahler moduli space we should certainly consider the associated
moduli fields, that is the vector multiplet scalars $b^i$. Clearly, we
should also keep the transition states $q^u$ which become light at the
flop. However, these states are charged and generically source the
vector fields. Hence, it seems we have to allow for non-trivial vector
field backgrounds for consistency. Fortunately, we can avoid such a
considerable complication by setting all scalars $q^u$ equal to each
other, that is, $q\equiv 2q^u$ for all $u=1,2,3,4$ and a single scalar
$q$. This configuration is consistent with the $q^u$ equations of
motion, as can be seen from~\eqref{Vspec}, and leads to a vanishing
current~\eqref{current}. Consequently, the vector fields can be
consistently set to zero in this case. The standard hypermultiplets,
in fact, completely decouple from the other fields and are, hence, not
essential for our purpose. From these standard hypermultiplet scalars
we will only keep the dilaton $\cV = e^\f$ since it represents the
overall volume of the internal Calabi-Yau space and is, therefore, of
particular physical relevance.

In summary, the spectrum of our five-dimensional effective action can
be consistently truncated to the five-dimensional metric, the
$h^{1,1}-1$ K\"ahler moduli space scalars $b^i$, the universal
transition scalar $q$ and the dilaton $\f$. From Eq.~\eqref{SV} and
Eq.~\eqref{SHspec}, the accordingly truncated effective action then
reads
\begin{eqnarray}\label{tr-action}
S_{5}&=&-\frac{1}{2\k_{5}^{2}}\int_{M_{5}}d^{5}x\sqrt{-g}\bigg\{
            \frac{1}{2}R+\frac{1}{4}\pt_{\a}\f\pt^{\a}\f
             +\frac{1}{4}G_{kl}\pt_{\a}b^{k}\pt^{\a}b^{l}+\lambda(K-6)\\
          &&\qquad\qquad\qquad\qquad\qquad\;
          +\pt_{\a}q\pt^{\a}q+V\bigg\}\, ,\nn\\
       V&=&\frac{1}{4}g^2
       \left[(\b_{l}b^{l})^{2}q^2
       +(G^{kl}\b_{k}\b_{l}-(\b_{l}b^{l})^{2})q^4\right]\; .
       \label{potential}
\end{eqnarray}
A Lagrange multiplier term has been added to enforce the constraint
$K=6$, Eq.~\eqref{cons}, on the moduli $b^{i}$. The value
of the gauge coupling $g$ has been given in Eq.~\eqref{g}. We also
recall that the ``K\"ahler potential'' $K$ and the metric $G_{ij}$
have been defined in Eq.~\eqref{K} and Eq.~\eqref{G}, respectively.

\vspace{0.4cm}

We are now ready to consider the cosmological evolution of our system.
We focus on backgrounds depending on time $\t$ only and a metric with
a three-dimensional maximally symmetric subspace which we take to
be flat, for simplicity. Accordingly, we consider the following 
Ansatz
\begin{alignat}{3}
  ds^{2}&=-e^{2\n (\t )}d\t^{2}+e^{2\a (\t)}d{\bf x}^2
          +e^{2\b (\t )}dy^{2}\,\\
  \phi & =\phi(\t)\\
  b^{i}&=b^{i}(\t)\\
  q&=q(\t)\; ,
\end{alignat}
where ${\bf x}=(x^1,x^2,x^3)$ and $y=x^4$. Note that $\a$ and $\b$
are the scale factors of the three-dimensional universe and the
additional spatial dimension, respectively. For later convenience, we have
also included a lapse function $\n$. The equations of motion for
this Ansatz, derived from the action~\eqref{tr-action}, are given
by
\begin{itemize}
 \item Einstein equations:
  \begin{eqnarray}
   3(\dot{\a}^{2}+\dot{\a}\dot{\b})&=&+\frac{1}{4}\left(\dot{\phi}^{2}
   +G_{ij}\dot{b}^{i}\dot{b}^{j}+4\dot{q}^{2}\right)+e^{2\n}\,V\nn
   \\
    3(\ddot{\a}-\dot{\n}\dot{\a}+2\dot{\a}^{2})&=&
    -\frac{1}{4}\left(\dot{\phi}^{2}+G_{ij}\dot{b}^{i}\dot{b}^{j}
    +4\dot{q}^{2}\right)+e^{2\n}\,V\label{Eineq}\\
  2\ddot{\a}+\ddot{\b}+3\dot{\a}^{2}+\dot{\b}^{2}+2\dot{\a}\dot{\b}
  -2\dot{\n}\dot{\a}-\dot{\n}\dot{\b}&=&-\frac{1}{4}\left(\dot{\phi}^{2}
  +G_{ij}\dot{b}^{i}\dot{b}^{j}
  +4\dot{q}^{2}\right)+e^{2\n}\,V\nn
  \end{eqnarray}
 \item Field equations of motion:
  \begin{eqnarray}
   &\ddot{\phi}+(3\dot{\a}+\dot{\b}-\dot{\n})\dot{\phi}=0&\\
   &\ddot{b}^{k}+(3\dot{\a}+\dot{\b}-\dot{\n})\dot{b}^{k}
     +\G_{ij}^{k}\dot{b}^{i}\dot{b}^{j}+2e^{2\n}\left( G^{kj}
    \frac{\pt V}{\pt b^{j}}-\frac{2}{3}b^iV\right) =0&\label{Fieldeq}\\
   &\ddot{q}+(3\dot{\a}+\dot{\b}-\dot{\n})\dot{q}
    +\frac{1}{2}e^{2\n}\,\frac{\pt V}{\pt q}=0&\nn\\
    &K=6\,.&\nn
  \end{eqnarray}
\end{itemize}
In these equations, we have already used the result $\lambda=-V/9$ for the
Lagrange multiplier $\lambda$ which follows by contracting the $b^i$
equations of motion~\footnote{Some relations for very special geometry
which are useful in this context have been collected in
Ref.~\cite{Lukas:1998tt}.} with $b_i=G_{ij}b^j$. The
connection $\G_{ij}^k$ associated to the moduli space metric $G_{ij}$
has been defined in Eq.~\eqref{Gamma}.

The above action and evolution
equations have been written for a definite topology of the internal
space with intersection numbers $d_{ijk}$. When an evolution leads
to a flop transition, the K\"ahler potential $K$, the metric $G_{ij}$
and the connection $\G_{ij}^k$ have to be changed "by hand" in
accordance with the change~\eqref{ichange} in the intersection numbers
to obtain the equations of motion for the new topology. As discussed
earlier, this implies continuity of $K$ and the metric while the
connection jumps across the transition. We also note that, from
Eq.~\eqref{potential}, the potential $V$ is continuous while its
derivatives with respect to $b^i$ contain the connection and, hence,
jump. From these properties and the equations of motion we conclude
that all fields and their first time derivatives and, hence, the
stress energy for all fields is continuous across the flop.   

\subsection{An approximate solution for vanishing transition states}
\label{solsect}

It is clear from the Eqs.~\eqref{Eineq} and \eqref{Fieldeq} that the
transition state $q$ can be set to zero consistently and we will now
analyse the cosmological evolution in this case. From the structure of
the equations of motion, the configuration $q=0$ seems rather
non-generic and having a non-vanishing, perhaps small initial value
for $q$ appears to be more plausible. We will study this generic case
further below. However, setting $q=0$ corresponds to the conventional
picture of a flop transition as being induced by slow free rolling
in moduli space. It is, therefore, useful to consider this case in
some detail, if only as a point of reference.

\vspace{0.4cm}

Setting the transition state $q$ and, hence, the potential $V$, to zero
simplifies the equations of motion considerably. Another simplification
arises if we choose the gauge $\n =3\a +\b$ for the lapse function.
In this gauge, we will denote time by $\t$ in the following.
The second term in the $b^i$ equations vanishes for this choice and
multiplying the remainder by $\dot{b}_i$ we find by integration that
\begin{equation}
   k\equiv G_{ij}\frac{\pt b^{i} }{\pt \t}\frac{\pt b^{j}}{\pt \t}
  =\mbox{const.}\,.
\end{equation}
Hence, the kinetic energy $k$ of the K\"ahler moduli is constant on
hyper-surfaces of constant time $\t$. Note that this is no longer
true for proper cosmological time related to $\t$ by
$dt^{2}=e^{6\a +2\b}d\t^{2}$. It is straightforward to integrate the Einstein
equations~\eqref{Eineq} and the $\f$ equation of motion for time $\t$.
The result can be easily rewritten in terms of proper time $t$ where it
takes the form
\begin{equation}
  \a=p_{\a}\ln|t|\,,\qquad \b=p_{\b}\ln|t|
  \,,\qquad
  \phi=p_{\phi}\ln|t|\, . \label{power}
\end{equation}
Here, we have dropped trivial additive integration constants for all
three fields and the origin of time, for simplicity. The expansion
powers $p_\a$, $p_\b$ and $p_\f$ must satisfy the constraints
\begin{eqnarray}
  3p_{\a}+p_{\b}&=&1\,,\label{p1}\\
  p_{\a}^{2}+p_{\a}p_{\b}&=&\frac{1}{12}\left(p_{\phi}^{2}
  +k\right)\,,\label{p2}
\end{eqnarray}
and the relation between proper time $t$ and $\t$ is simply
\begin{equation}
 \t = \ln |t|\; .
\end{equation}
We still need to find the explicit form of $b^i$ for a complete solution.
To do this, we have to solve the following system of equations
\begin{eqnarray}
 \ddot{b}^i+\G_{jk}^i\dot{b}^j\dot{b}^k &=& 0 \label{geo}\\
 d_{ijk}b^ib^jb^k &=& 6 \label{c1}\\
 G_{ij}\dot{b}^i\dot{b}^j &=& k\; .\label{c2}
\end{eqnarray}
Here the dot denotes the derivative with respect to $\t$. Hence, in this
time coordinate, the fields $b^i$ move along geodesics in moduli space
subject to the constraint~\eqref{c1} from special geometry and the
kinetic energy constraint~\eqref{c2}. Unfortunately, the equation~\eqref{geo}
is hard to solve in general due to the second non-linear term. 
However, for a sufficiently small time interval and slow motion this
term can be neglected. In other words, the geodesics are well approximated
by straight lines
\begin{equation}
 b^i = c^i+p^i\t + O(\t^2)\; \label{blin}
\end{equation}  
in moduli space, where $c^i$ and $p^i$ are constants, as long as
\begin{equation}
 |\t |\ll \left|
    \frac{2p^{i}}{\G^{i}_{jk}(c)p^{i}p^{j}}\right|\, .\label{validity}
\end{equation}
This approximation also implies that we neglect the kinetic energy of
the fields $b^i$ compared to the other fields, that is, from
Eq.~\eqref{c2}, we consider a solution with $k\simeq 0$. Accordingly,
this value for $k$ has to be inserted into the
relation~\eqref{p2}. Within our approximation, the special geometry
constraint~\eqref{c1} turns into two conditions, namely
\begin{equation}
 d_{ijk}c^ic^jc^k = 6\; ,\qquad d_{ijk}c^ic^jp^k=0\; .\label{c3}
\end{equation}
These algebraic equations can be easily solved for given intersection numbers.
This completes our approximate solution.

\vspace{0.4cm}

Let us now apply this result to a flop transition. We assume, for
simplicity, that the flop occurs in the $b^1$ direction and at
$b^1=0$. By setting $c^1=0$ in our solution~\eqref{blin} we can, in
fact, arrange the flop to take place at time $\t = 0$. Now we consider
two solutions of the above type with intersection numbers $d_{ijk}$
and $\tilde{d}_{ijk}$. We recall that $d_{111}$ is the only
intersection number which changes (as given in Eq.~\eqref{ichange})
across the transition. Since we have set $c^1=0$ this particular
intersection number drops out of the constraints~\eqref{c3} which need
to be satisfied for a valid solution. We are, therefore, free to
choose the same constants $c^i$, $p^i$, solving the constraints~\eqref{blin},
on both sides of the flop. This leads to two solutions, for either
topology, which can be continuously matched together at the flop
transition for $\t =0$. Further, our approximation is valid for
a certain period of time before and after the flop as quantified
by the condition~\eqref{validity}. The scale factors $\a$ and $\b$ and
the dilaton $\f$ are unaffected by the transition in that they evolve
according to~\eqref{power} with the same expansion powers $p_\a$,
$p_\b$ and $p_\f$ on both sides of the transition. These results
suggest that the system indeed evolves through the transition into the
moduli space of the topologically distinct Calabi-Yau space and,
hence, that the topology change is dynamically realized. This picture
will be confirmed by our numerical integration further below.

\subsection{Evolution for non-vanishing transition states}

If the transition states no longer vanish, that is $q\neq 0$, the
potential becomes operative and the conclusions of
the previous subsection do not apply any more. Clearly, we should not
expect to find analytic solutions in this case any more. However, some
qualitative features of the evolution can be inferred from the
structure of the potential~\eqref{potential}.

Let us consider small values of the transition state such that the
potential~\eqref{potential} is dominated by the first term.  It is
then approximately given by $V\sim b^2q^2$. Note, that this potential,
for fixed non-zero $q$ has a minimum at $b=0$, that is, precisely at
the flop point. From this observation and the general shape of the
potential it is intuitively clear that a generic evolution will lead
to oscillations around $b=0$ and will finally settle down to this
point. In other words, there is a clear preference for the system to
settle down at the transition point rather than complete the
transition.

The complete potential~\eqref{potential} must still have a minimum at
small $b$ for fixed non-zero $q$ at least as long as $q$ is
sufficiently small. This suggests that the above argument generalises
to this case and this will indeed be confirmed by our numerical
integration. In conclusion, this suggest that the system behaves quite
differently if we allow non-zero values of the transition states.
Previously, for vanishing transition states, we have found that
the topology does change dynamically. If $q\neq 0$, on the other hand,
the potential becomes important and favours the region in moduli
space close to the flop transition. In this case, the system tends
to settle down in the transition region so that the topology
change is not completed.


\section{An explicit model}\label{example}

In this section, we would like to substantiate our previous claims by
numerically studying the cosmological evolution of our system. To do
this we need to consider a particular example, that is, a particular
pair of Calabi-Yau manifolds related by a flop transition which
provides us with a concrete set of intersection numbers. We will use
the Calabi-Yau spaces described in
Refs.~\cite{Louis:1996mt,Morrison:1996pp,Candelas:1994hw} and applied
to black hole physics in Refs.~\cite{Chou:1997ba,Gaida:1998km}.

Concretely, we consider two elliptically fibred Calabi-Yau spaces $X$
and $\tilde{X}$, both with a Hirzebruch $F_1$ base space and with
$h^{1,1}=3$. These spaces share a boundary of the K\"ahler moduli
space which corresponds to a flop transition. Following
Ref.~\cite{Chou:1997ba}, both moduli spaces can be covered by a single set of
coordinates $(W,U,T)$ with the flop transition along $W=U$.
The K\"ahler moduli space of $X$ corresponds to the coordinate range
\begin{equation}
U>W>0\; ,\qquad T>\frac{3}{2}U\, , \label{r1}
\end{equation}
and the associated K\"ahler potential is given by
\begin{equation}
 K=\frac{9}{4}U^{3}+3T^{2}U-W^{3}\; . \label{k1}
\end{equation}
We can use the constraint $K=6$ to solve for $T$ in terms of the
other two moduli resulting in
\begin{equation}
 T=\frac{1}{6}\left(-\frac{27U^{3}-12W^{3}-72}{U}\right)^{1/2}\; .
\end{equation}
The fields $b^i$ which we have previously used are related to
$(W,U,T)$ by
\begin{equation}
  b^{1}=U-W\; ,\qquad b^{2}=W\; ,\qquad
  b^{3}=T-\frac{3}{2}U\; .
\end{equation}
This definition implies, from Eq.~\eqref{r1}, that the fields $b^i$
are indeed positive throughout the moduli space of $X$. The flop
transition is approached as $b^1\rightarrow 0$. Hence, the
coefficients $(\b^i)$ which enter the potential~\eqref{potential}
are given by $(\b^1,\b^2,\b^3)=(1,0,0)$.
 
For the second Calabi-Yau space $\tilde{X}$ the moduli are in the range
\begin{equation}
 W>U>0,\qquad T>W+\frac{1}{2}U\, , \label{r2}
\end{equation}
and the K\"ahler potential is given by
\begin{equation}
 \tilde{K}=\frac{5}{4}U^{3}+3U^{2}W-3UW^{2}+3T^{2}U\; .\label{k2}
\end{equation}
Solving $\tilde{K}=6$ for $T$, as before, leads to
\begin{equation}
 \tilde{T}=\frac{1}{6}\left(-\frac{15U^{3}+36U^{2}W-36UW^{2}-72}{U}
           \right)^{1/2}\; .
\end{equation}
From Eq.~\eqref{bt}, fields $\tilde{b}^i$ which are positive
in the moduli space of $\tilde{X}$ can be defined by
\begin{equation}
  \tilde{b}^{1}=-b^1=W-U\; ,\qquad \tilde{b}^{2}=b^2-b^1=2W-U\; ,\qquad
  \tilde{b}^{3}=b^3-b^1=T-\frac{5}{2}U+W\; .
\end{equation} 
Note, that the two moduli spaces~\eqref{r1} and \eqref{r2}
indeed have a common boundary at $b^1=U-W=0$ which is where the
flop transition occurs. Also, using the basis $b^i$ as defined above,
it is easy to see that the K\"ahler potentials~\eqref{k1} and
\eqref{k2} are related by the shift~\eqref{ichange} in the intersection
numbers, as is required for a flop transition. 

\vspace{0.4cm}

\begin{figure}
\begin{center}
\begin{tabular}{cc}
  \includegraphics[scale=0.8]{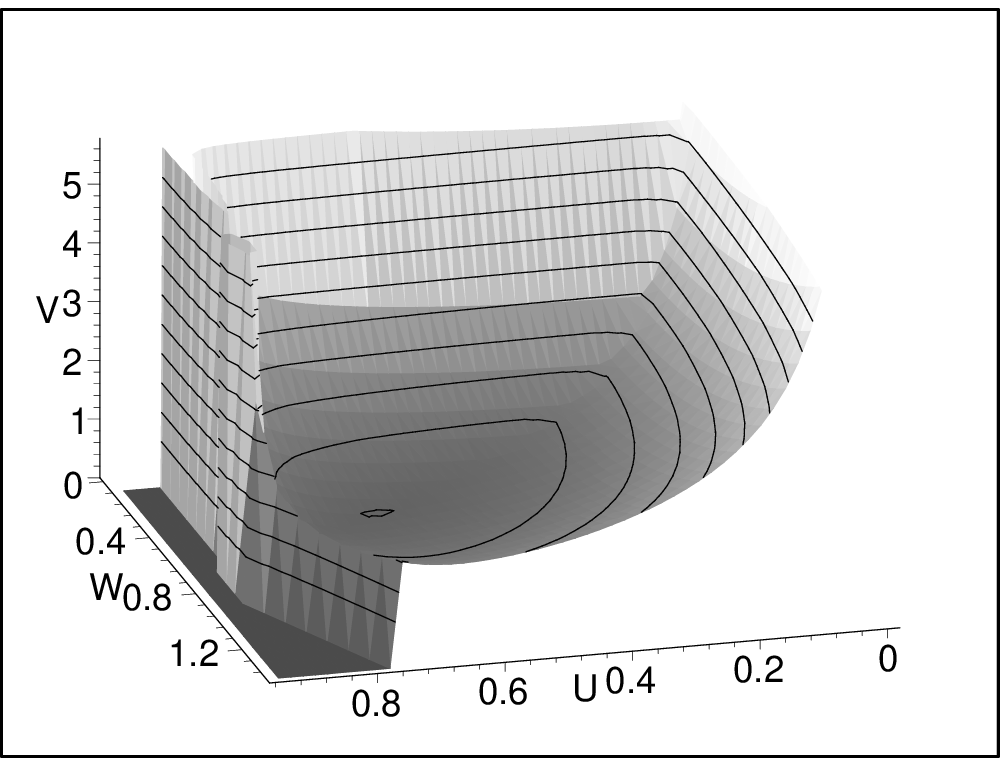}&
  \includegraphics[scale=0.8]{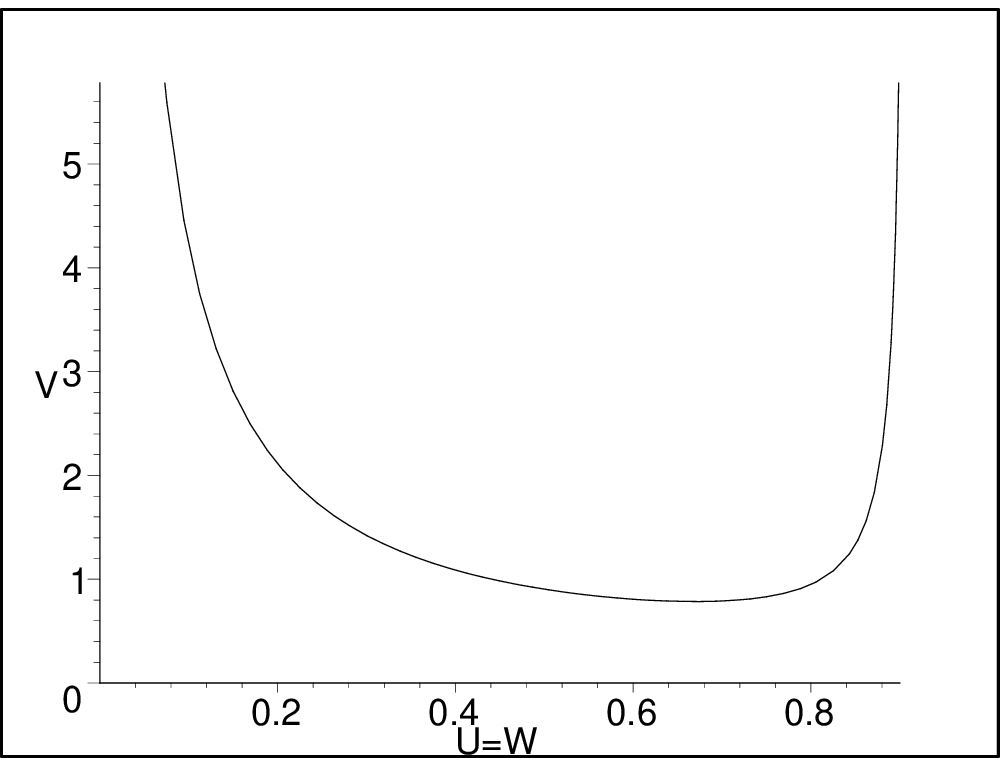}
\end{tabular}
  \caption{Potential V in terms of W,U and along the flop transition
           line at W=U, both for fixed q=1/3.}
  \label{fig:potential}
\end{center}
\end{figure}

We will now study the above example using $W$ and $U$ or, equivalently,
$b^1$ and $b^2$ as the independent variables. It is useful to plot
the potential~\eqref{potential} as a function of these variables for
a fixed value of $q$. This has been done in Fig.~\ref{fig:potential}
for a value of $q=1/3$. 
Obviously, this potential has a minimum in both directions which happens
to be at
\begin{equation}
 U=W=\left(\frac{3}{10}\right)^{1/3}\label{UWmin}
\end{equation}
independent of the value of $q$. The associated potential value
at the minimum (in units where $g=1$) is
\begin{equation}
 V_{\rm min} = \frac{300^{2/3}}{16}q^4\; .\label{Vmin}
\end{equation}
From our previous general argument, we did anticipate a minimum in the
direction $b^1=U-W$. However, a minimum for both fields, located
precisely at the flop and, hence, at field values~\eqref{UWmin}
independent of $q$, comes as a surprise. We do not know whether this
is a general feature of the potential~\eqref{potential} near flop
transitions or particular to this example. Incidentally, we note that,
having fixed all moduli $b^i$, the shape of the potential~\eqref{Vmin}
in the remaining $q$-direction is well-suited for
inflation. Unfortunately, to be in the slow-roll regime we need that
$q\gg 1$ which, in turn, implies that $V\gg 1$ in units of the
fundamental Planck scale. Such potential values clearly go beyond the
region of validity of our five-dimensional effective action and it is,
therefore, difficult to conclude anything definite about this
tantalising possibility. In any case, we will restrict field values
such that $V$ does not become too large in the following.

Before studying the flop-transition numerically, let us briefly discuss
the other boundaries of the moduli space as described by the
conditions~\eqref{r1} and \eqref{r2} in relation to the potential.
We note that the potential steeply increases towards the boundary
directions $W\rightarrow 0$, $W\rightarrow\infty$ and $U\rightarrow 0$.
Hence, as long as the potential is operative (that is, $q$ is non-zero)
it prevents evolution towards these boundaries. The same is true
for the boundary prescribed by $U^3\rightarrow (W^3+6)/9$ 
as long as one stays in the $X$ part of the moduli space, $b^1=U-W>0$,
and $W$ is sufficiently small. The potential barrier rapidly vanishes
in this direction for increasing $W$ in the $\tilde{X}$ part of the
moduli space. In our numerical evolution, we will simply avoid this
direction of moduli space by choosing suitable initial conditions. 

\vspace{0.4cm}

\begin{figure}[ht]
\begin{center}
\begin{tabular}{ccc}
  \includegraphics[scale=0.52]{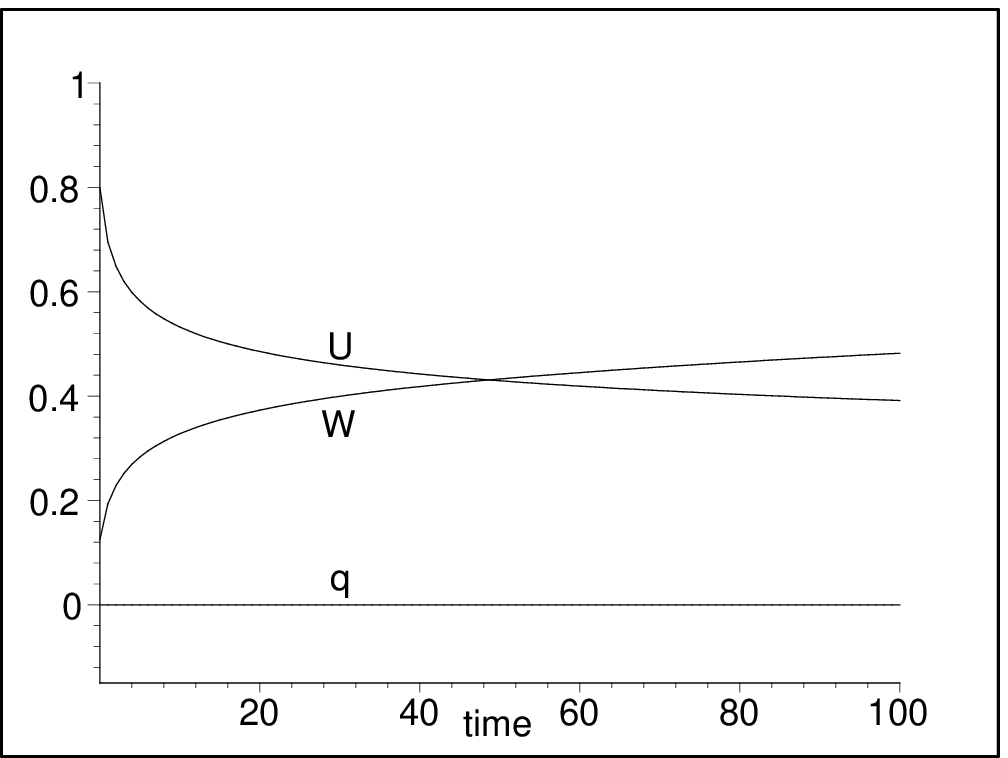}&
  \includegraphics[scale=0.52]{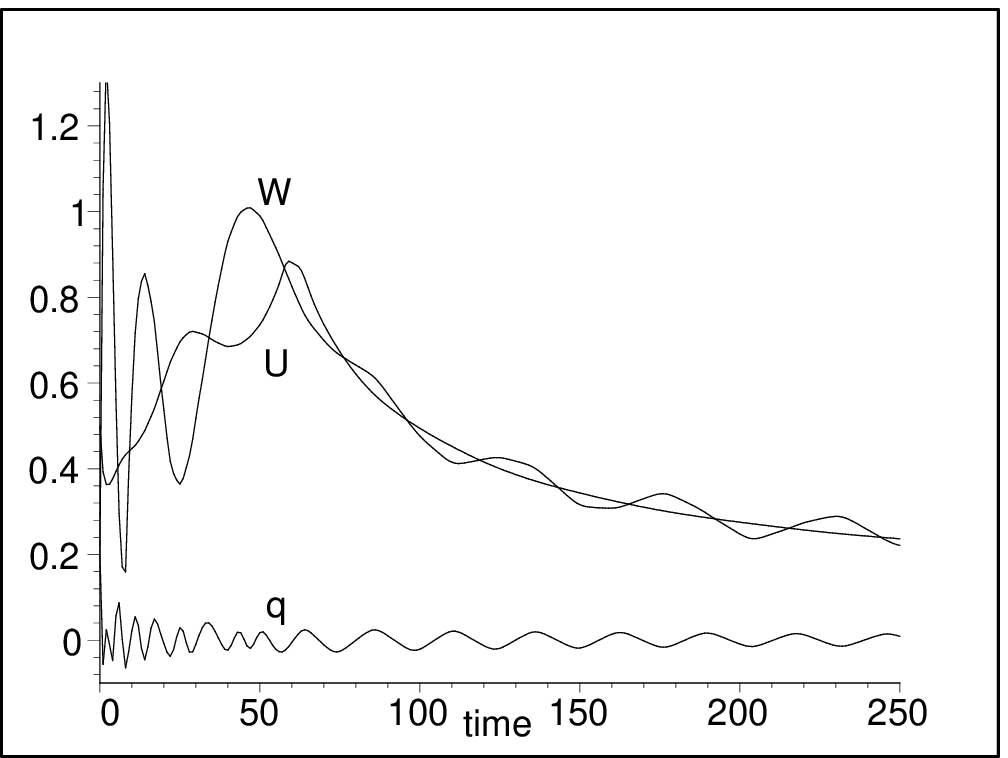}&
  \includegraphics[scale=0.52]{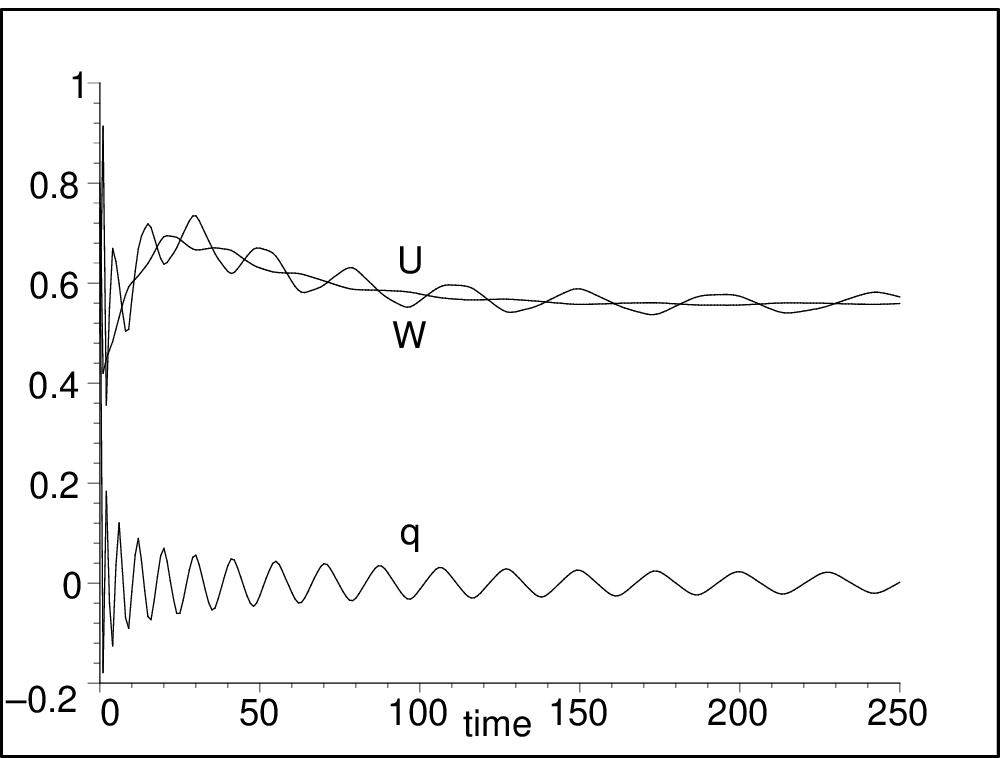}\\
  \includegraphics[scale=0.52]{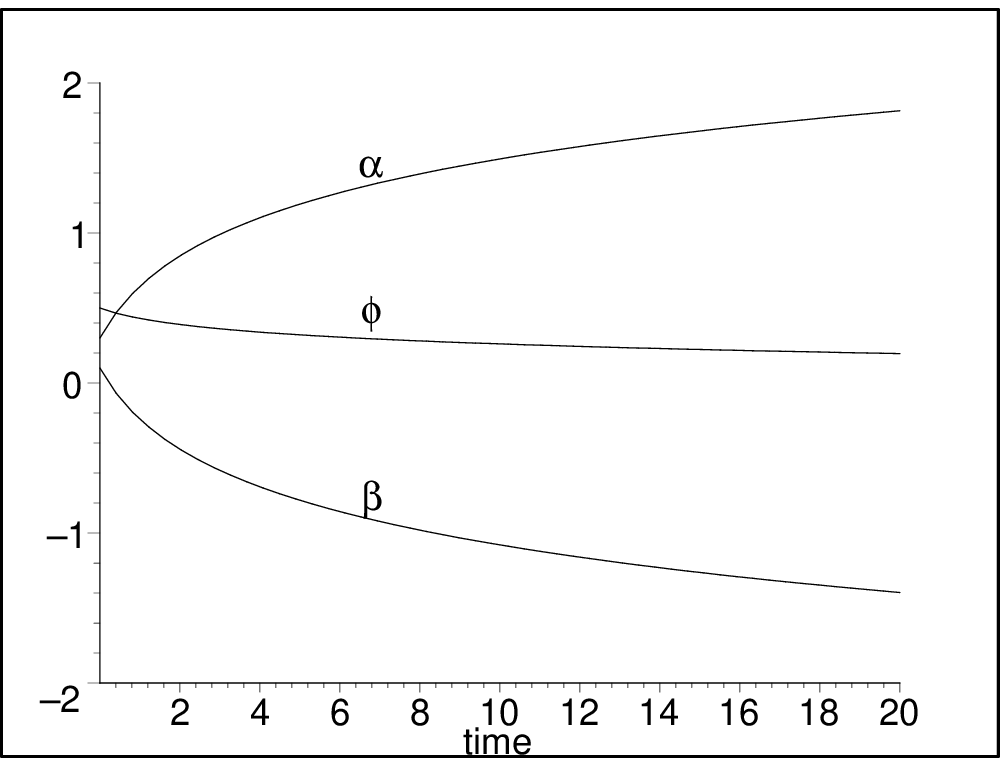}&
  \includegraphics[scale=0.52]{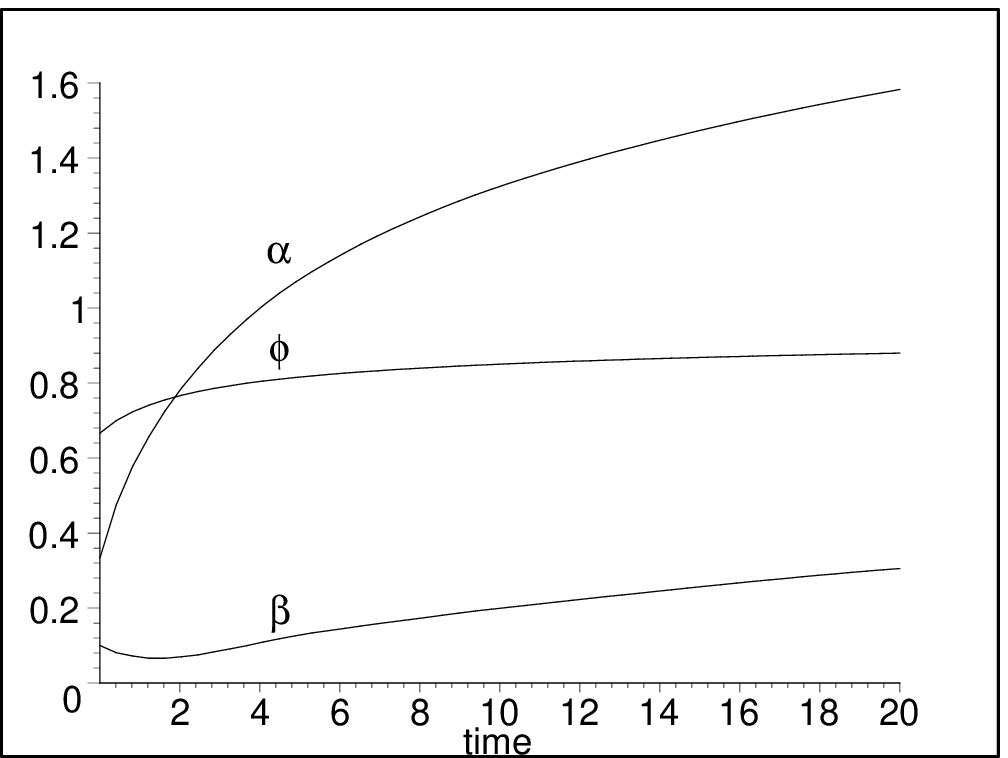}&
  \includegraphics[scale=0.52]{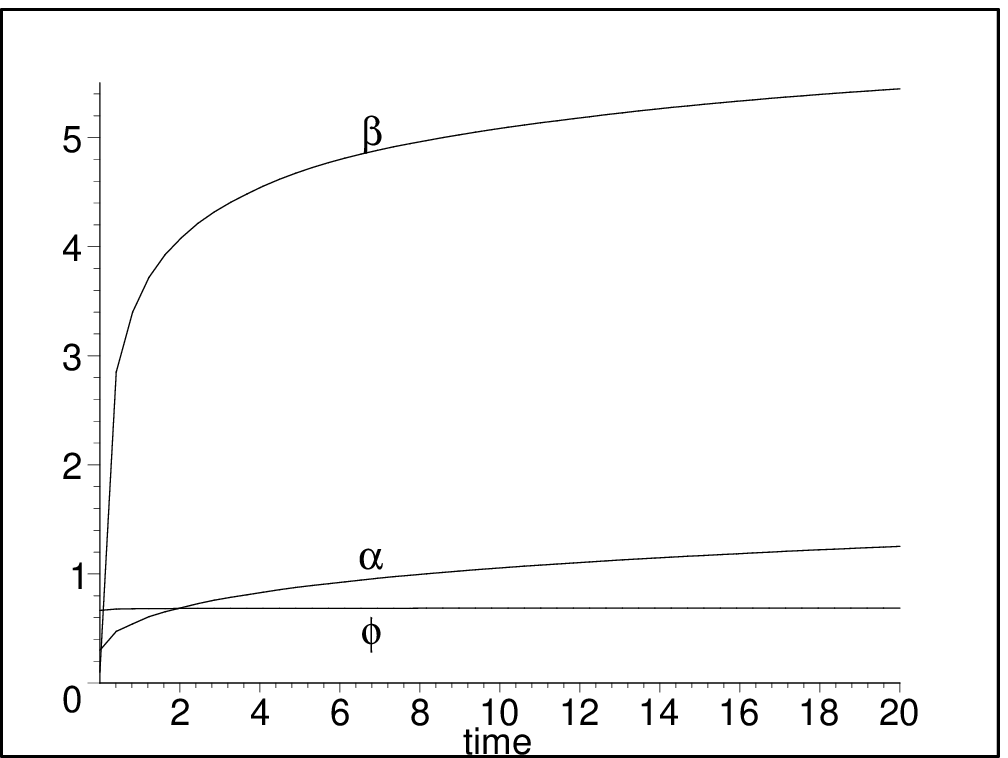}
\end{tabular}
\caption{Cosmological behaviour of the fields W(t), U(t), q(t) and
$\a(t)$, $\b(t)$, $\phi(t)$, respectively, for the three different sets of
initial conditions given in table \ref{tab:inicond}.}
\label{fig:plots}
\end{center}
\end{figure}

We have numerically integrated the system of equations~\eqref{Eineq},
\eqref{Fieldeq} for the above example, that is for the K\"ahler
potentials~\eqref{k1} and \eqref{k2}. Here, we present the results
for three characteristic sets of initial conditions which lead to
an evolution towards the flop transition region. The precise
initial values of all fields are specified in table~\ref{tab:inicond}.
\begin{table}[b]
\begin{center}
\begin{tabular}{|c||c|c|c|c|c|c|c|c|c|c|c|}
 \hline
 plot & $U$ & $\dot{U}$ & $W$ & $\dot{W}$ & $q$ & $\dot{q}$ &
 $\phi$ & $\dot{\phi}$ & $\a$ & $\dot{\a}$ & $\b$ \\ \hline\hline
 $1^{st}$ & 4/5 & -1/5 & 1/8 & 1/9 & 0 & 0 & 1/2 & -1/10 & 3/10 & 1/2 &
 1/10 \\ \hline
 $2^{nd}$ & 1/2 & -1/5 & 1/5 & 1/10 & 1/5 & 1/8 & 2/3 & 1/10 & 1/3 & 2/5 &
 1/10 \\ \hline
 $3^{rd}$ & 4/5 & -3/10 & 1/2 & 1/10 & 3/4 & 1/9 & 2/3 & 1/5 & 3/10 & 3/10 &
 1/10 \\ \hline
\end{tabular}
\caption{Table of initial conditions in order of increasing
initial value of $q=q(0)$.}
 \label{tab:inicond}
\end{center}
\end{table}
Fig.~\ref{fig:plots} shows the corresponding evolution of the fields
as a function of proper time $t$. The first set of initial conditions
in table~\ref{tab:inicond} leads to a vanishing transitions state,
that is, we have chosen $q(0)=0$ and $\dot{q}(0)=0$. This is precisely
the case we have discussed in subsection~\eqref{solsect}. The
resulting evolution is shown in the first column of
fig.~\ref{fig:plots}. It can be seen that the system, starting off in
the moduli space of $X$ at $b^1>0$, evolves towards the flop
transition $b^1=U-W\rightarrow 0$ and then moves on to negative values
of $b^1$, corresponding to the moduli space of $\tilde{X}$. Hence, the
topological transition is indeed dynamically realized as suggested by
the previous analytic solution. The picture changes considerably once
we allow for a non-vanishing transition state. The second set of
initial conditions in table~\ref{tab:inicond} corresponds to small,
non-vanishing values of $q(0)$ and $\dot{q}(0)$. Again, the system
starts out in the $X$ moduli space and evolves towards the flop. The
associated plots in the second row of fig.~\ref{fig:plots} show that
after a few large initial oscillations around $b^1=0$ the system
stabilises at $b^1=U-W\simeq 0$ and, hence, at the flop transition. A
similar behaviour can be observed for larger initial values $q(0)$, as
in the third set in table~\ref{tab:inicond} with associated plots in
the third column of fig.~\ref{fig:plots}.


\section{Conclusions}

We have shown in this paper that the dynamics of M-theory flop
transitions strongly depends on whether or not the transition states
which become light at the flop are taken into account. If these modes
are exactly set to zero the moduli space evolution proceeds freely and
the topology change can indeed be dynamically realized, that is, the
system moves between two topologically different Calabi-Yau spaces
related by a flop for appropriate but generic initial conditions.  For
non-vanishing values of the transition modes, however, a potential
becomes operative which generically stabilises moduli fields at the
flop. Hence, the system does not really evolve into the moduli space
of the flopped Calabi-Yau manifold and the transition remains
incomplete. One may argue that this latter case is likely since
non-vanishing values of the transition states represent a more generic
set of field configurations in the early universe. If this is indeed
the case, the region in moduli space close to a flop is preferred by
the dynamics of the system.

It is likely that our results can be transferred to heterotic M-theory
which provides a more realistic setting for low-energy physics from
M-theory. To do this, we have to compactify the fifth dimension on a
line interval and couple the five-dimensional $N=1$ bulk supergravity
used in this paper to $N=1$ theories on the two
boundaries~\cite{Lukas:1998yy,Lukas:1998tt}. The vacuum state of this
theory is a static BPS domain wall which corresponds to a certain path
in the Calabi-Yau K\"ahler moduli space as one moves between the
boundaries. Using this property of the vacuum state, one can, in fact,
construct static vacua of heterotic M-theory with an inherent flop
transition~\cite{Greene:2000yb}, that is, vacua with the flop occurring
at a particular point in the interval. Assuming the results of this
paper indeed transfer to heterotic M-theory, it is these inherently
flopped vacua which would be preferred by the dynamical evolution of
the system.


\vspace{1cm}

\noindent
{\Large\bf Acknowledgements}\\
We would like to thank Natxo Alonso-Alberca for collaboration at an
early stage of this project. A.~L.~is supported by a PPARC Advanced
Fellowship. M.~B.~is supported by a Graduiertenkolleg Fellowship of
the DFG.


\end{document}